# A Peregrine soliton-like structure that has nothing to deal with a Peregrine breather

**Christophe Finot**

Laboratoire Interdisciplinaire Carnot de Bourgogne, UMR 6303 CNRS-Université Bourgogne-Franche Comté, 21000 Dijon, France

**\* Correspondence:**
Corresponding Author
christophe.finot@u-bourgogne.fr

**Abstract**

We report on experimental results where a temporal intensity profile presenting some of the main signatures of the Peregrine soliton (PS) is observed. However, the emergence of a highly peaked structure over a continuous background in a normally dispersive fiber cannot be linked to any PS dynamics and is mainly ascribed to the impact of Brillouin backscattering.

## 1    Introduction

Peregrine soliton (PS) or breather that was initially analytically derived by H. Peregrine [1] has since its first experimental demonstration in a physical system in 2010 [2] generated a very large number of experimental and theoretical studies driven primarily by the search for extreme events [3]. This coherent structure presenting temporal and spatial localizations has therefore been extensively investigated first in the field of fiber optics [2, 4] and then in the hydrodynamic domain [5] and multicomponent plasmas [6]. In optics, PS is a rational solution of the nonlinear Schrodinger equation (NLSE) that describes the evolution of light in a fiber with focusing nonlinearity: the combination of anomalous dispersion and self-phase modulation induced by Kerr nonlinearity leads to the emergence of a wave that appears from nowhere and disappears without a leaving trace [3]. The list of the fiber-based nonlinear processes that are linked to the PS evolution is a long one and includes the modulation instability [7], the propagation of turbulent fields [8-10], the initial stage of temporal compression of higher-order soliton pulses [11], the focusing of super-Gaussian structures [12], the generation of ultrashort structures [13]… Peregrine-like structures have also more recently been observed in other photonic systems that may deviate from the simple NLSE framework. Indeed, PS or breather features have been experimentally or theoretically reported in lasers [14], optical cavities [15], metamaterials [16], quadratic or photorefractive media [17, 18].

In this present brief report, we discuss a simple experimental fiber-based configuration where temporal intensity profiles presenting several signatures of the PS waveform are recorded. However, we demonstrate that even if intriguing similarities may exist, the observed coherent structure must be carefully analyzed and we conclude that our observations should not be straightforwardly associated with a PS dynamics.

## 2    Method

The experimental setup we implemented is sketched in Fig. 1A and relies on devices from the telecommunication industry that are commercially available. A continuous wave at 1550 nm is first intensity modulated thanks to a lithium niobate modulator operating at its point of maximum transmission and driven by an electrical pulse generator that delivers super-Gaussian pulses at a repetition rate of 2 GHz. The resulting temporal intensity profile is plotted in Fig. 1B and is a continuous wave where light has been switched off for a duration of 40 ps. Particular care has been devoted to the optimization of the extinction ratio in order to prevent parasitic interference between a residual unwanted background and the main structures [19]. The temporal profile can be well fitted by an inverted second-order super-Gaussian pulse. Its spectrum is recorded on a high-resolution optical spectrum analyzer (see Fig. 1C) and exhibits a high signal-to-noise ratio as well as a high level of symmetry and coherence degree (the spectral linewidth of the components being below the resolution of the optical spectral analyzer, i.e. 5 MHz). The signal is then amplified thanks to a low-noise erbium-doped fiber amplifier that delivers an average power that can be continuously tuned from 10 dBm up to 23 dBm. Propagation occurs in a single spool of 10.5 km of dispersion-shifted fiber with normal dispersion (second-order dispersion $\beta_2$ of 19 ps$^2$/km and a nonlinear coefficient $\gamma$ of 2 /W/km$^{-1}$). After propagation, the signal is analyzed with a photodiode and a high-speed sampling oscilloscope (electrical bandwidth of 50 GHz) as well as with an optical sampling oscilloscope.

Light propagation in a single-mode fiber can be described by the NLSE that also includes losses. The evolution of the complex scalar field $\psi(t,z)$ in the slowly varying approximation is therefore governed by [20] :

$$i\frac{\partial \psi}{\partial z} = \frac{1}{2}\beta_2 \frac{\partial^2 \psi}{\partial t^2} - \gamma |\psi|^2 \psi - \frac{\alpha}{2}\psi \quad (1)$$

with $z$ and $t$ being the longitudinal and temporal coordinates, respectively. $\alpha$ is a coefficient of optical losses. This equation can be numerically solved by the widely used split-step Fourier algorithm [20]. When light propagates in an anomalous dispersive fiber, one of the solutions of the loss-free NLSE is the PS which temporal profile $\psi_S(\tau)$ at the point of maximum focusing is described by the following typical rational solution :

$$\psi_S(\tau) = 1 - \frac{4}{1+4\tau^2} \quad (2)$$

where $\psi_S$ is the field normalized with respect to the continuous background and $\tau$ a normalized time depending on the system parameters.

## 3    Results

The temporal output profile recorded on a high-speed sampling oscilloscope for the highest input power available is plotted in Fig. 2A (solid blue line). For this power, the initial 40 ps dip of light has significantly reshaped into a highly peaked structure having a temporal duration at half-maximum of 16.6 ps. The highly symmetric intensity profile goes down to a minimum value at $t = \pm 28$ ps and lies on a continuous background that is 9 times weaker than the peak intensity. Results recorded independently on an optical sampling oscilloscope (Fig. 2B) fully confirm these features of the intensity profile and stress that the minimum value of the wave is close to zero. The overall profile can be adjusted by the typical temporal intensity waveform of the PS at the point of maximum compression (Eq. (2), with $\tau$ adjusted to fit the experimental data, red dotted lines). Such a similarity is deeply intriguing as PS are supposed to exclusively exist in presence of focusing nonlinearity, i.e. in the anomalous regime of dispersion. The evolution of the output profile according to the input power is summarized in Fig. 2C. It confirms that the ratio $R$ of the peak-power / continuous background continuously increases with the input power, similarly to what can be usually observed with breathers in the initial stage of the growth and decay cycle.

In order to confirm the experimentally recorded dynamics, we have carried numerical simulations based on the NLSE (Eq. (1)) including linear constant losses of 0.25 dB/km. The evolution of the output intensity profiles according to the input power is summarized in Fig. 3A. It is clear that NLSE in its simplest form completely fails to reproduce the main experimental findings. A moderate bump is visible at the center of the waveform and can be interpreted as the temporal analog of the Arago spot as we have detailed in [21]. However, its amplitude is well below the amplitude of the continuous wave. Even if its level increases with input power, numerical simulations ran for input powers up to 26 dBm do not predict any $R$ above 1. This inability of the usual NLSE model to provide even qualitative insights on the pulse dynamics forces us to reconsider the validity of the various assumptions made and to rethink our model.

## 4    Discussion

In order to better understand the physical origin of the major discrepancy between the numerical predictions and the experimental data, we recorded the output average power according to the input power. The results are reported in Fig. 3B and outline that the assumption of a constant level of loss severely fails. Indeed, whereas the input power is increased by one order of magnitude, the output power remains more or less constant. This indicates that the losses grow continuously with power. Those extra-losses are ascribed to Brillouin backscattering [20, 22] induced by the constant background. Note that in our previous experiments, we have conveniently mitigated this unwanted Brillouin by using a series of optical isolators [23] or an extra phase modulation [21]. Connections between Brillouin scattering and rogue events have already been highlighted in fiber lasers [24, 25], but to our knowledge never in the case of cavity-free passive propagation. In order to take into account these extra losses induced on the central coherent component, we have included a frequency and power dependence in the term of optical losses. More precisely, instead of complete and accurate modeling of the interactions of the signal with the Brillouin wave, we consider, as a very basic model, that, for the range of powers of interest, the losses can be described as:

$$\alpha_{dB}(\omega, Pin) = \alpha_{0,dB} + \delta(\omega)\left(A\,P_{in,dB} + B\right) \tag{3}$$

where $\alpha_0$ is the level of linear losses, $\omega$ the angular frequency and $A$, $B$ coefficients extracted from the experimental measurements. Note that our goal in this brief report is not to perform a detailed and accurate study of the Brillouin gain properties of fiber under investigation, but rather to identify the key elements that qualitatively explain our dynamics.

Results obtained with this new model are reported in Fig. 4A where we can make out the major differences compared to Fig. 3A. A peak emerges from the central part and the ratio $R$ between the peak power of the central structure and the continuous background is now clearly well above 1. The various trends recorded experimentally (Fig. 2C) become well captured by the numerical simulations. The inset of Fig. 4(A) shows results of the same simulations run with parameters of a virtual fiber having a dispersion coefficient opposite to the normally dispersive fiber under investigation ($\beta_2$ of -19 ps$^2$/km, all other parameters being kept identical). In this case, the central structure is much less pronounced, in agreement with the impact of the nonlinearity on the evolution of the temporal Arago spot [21].

Details of the temporal profile simulated in the normal regime of dispersion are provided in Fig. 4B for an input average power of 23 dBm and confirm the ratio of 9 between the peak intensity and the continuous background. Once again, the overall temporal intensity profile can be well fitted with the PS typical waveform and it is found that the intensity goes down to zero. The numerical simulations also enable us to get access to the phase profile (Fig. 4B2). Quite surprisingly, the observed phase profile also presents similarities with the phase profile of a PS that is characterized by a phase jump of π at the minimum of the intensity profile [26]. Profile simulated for propagation in the anomalous profile is also included and show that the central peak is surrounded by two non-negligible oscillations.

Note however that the physics involved in the emergence of the spiky structure is extremely different from the one involved in the PS dynamics. Contrary to the PS case where the wave emerges from the interaction of self-phase modulation and anomalous dispersion, the crucial component in our study is Brillouin scattering that depletes the continuous background and consequently increases the ratio $R$. In this context, normalization by the value of the continuous background may distort the interpretation as it gives the feeling that a strong peak emerges from an energy focusing process whereas the main effect is the drop of the value of the continuous background. One may also note that the coherent structure under investigation will not experience the growth-and-decay cycle typical of the PS and values of $R$ above 9 can be recorded as observed in panel 4A with $R = 12$ for $P_0 = 400$ mW. We can finally notice that contrary to the usual PS which temporal width decreases in the stage of temporal compression, both in experiments as well as in the numerics, the temporal duration of the central peak is not severely influenced by the input power, confirming that the underlying dynamics is very different.

## 5    Conclusions

To conclude, we have described an experimental configuration where several features of the Peregrine soliton seem to be reproduced during the propagation of a temporal hole of light in a fiber with normal dispersion. However, we demonstrate that despite these observed signatures, the physics that is involved is radically different and is essentially ascribed to the Brillouin backscattering that depletes the continuous background. With this example as well as another work dealing with breathers' features [27], we stress that great care should always be devoted when trying to identify the nature of coherent structures in an experimental record. A deep understanding of the underlying physical dynamics is required before claiming that extreme structures such as Peregrine solitons are observed in a system. In this context, numerical simulations are of great help. For the problem under investigation, whereas the standard NLSE was unable to retrieve the experimental features, adding nonlinear losses that only affect the continuous component was sufficient to qualitatively reproduce the influence of the input power on the output field. Finally, note that in a recent contribution, we have numerically stressed that PS could be observed a similar configuration when the intensity modulation is replaced by a phase modulation [28].

**Potential conflict of interest**

The authors declare that the research was conducted in the absence of any commercial or financial relationships that could be construed as a potential conflict of interest.

**Funding**

We acknowledge the financial support of the Institut Universitaire de France (IUF) and the Bourgogne-Franche Comté Region.

**Acknowledgement**

The research work has benefited from the PICASSO experimental platform of the University of Burgundy.

**Figure captions**

Fig 1 : (A) Experimental setup : CW : continuous wave ; IM : intensity modulator ; EDFA : erbium-doped fiber amplifier ; OSA : optical sampling oscilloscope ; PD : photodiode ; ESO : electrical sampling oscilloscope ; HR OSA : high-resolution optical spectrum analyzer. (B) Temporal intensity profile of the input pulse. Experimental results (circles) are fitted with an inverted super-Gaussian pulse (blue solid line) and compared with an ideal rectangular waveform (red). (C) Experimental optical spectrum of the input pulse.

Fig 2 : (A) Experimental output temporal profile recorded on the ESO (blue line) compared with a fit by the PS waveform (Eq. 2, red dotted line). Input average power is 23 dBm. (B) Same as (A) but recorded on the OSO. (C) Evolution of the temporal intensity profile according to the input power. Intensity profiles are normalized by the intensity of the output continuous background. The white dashed line indicates the temporal location of the sharp edges of the input waveform.

Fig 3 : (A) Numerical simulations of the output temporal intensity profiles based on the NLSE including constant losses. Intensity profiles are normalized by the intensity of the output continuous background. (B) Experimental measurement of the output average power (circles) according to the input average power. Results without any Brillouin mitigation (red circles) are compared with results

when mitigation is applied (blue circles). Diamonds represent the level of losses that is experienced: linear losses (blue diamonds) and nonlinear losses (red diamonds).

Fig 4 : (A) Numerical simulations of the output temporal intensity profiles based on the NLSE including power-dependent losses. Intensity profiles are normalized by the intensity of the output continuous background. Inset shows the results obtained in a virtual fiber with opposite dispersion. (B). Details of the temporal intensity and phase profiles (panel B1 and B2 respectively) obtained for an input power of 23 dBm: numerical simulations in the fiber with normal dispersion (blue line) are compared with a fit with a PS waveform (dotted red line). Blue dotted line represents the results predicted with a virtual anomalously dispersive fiber.

Figure 1

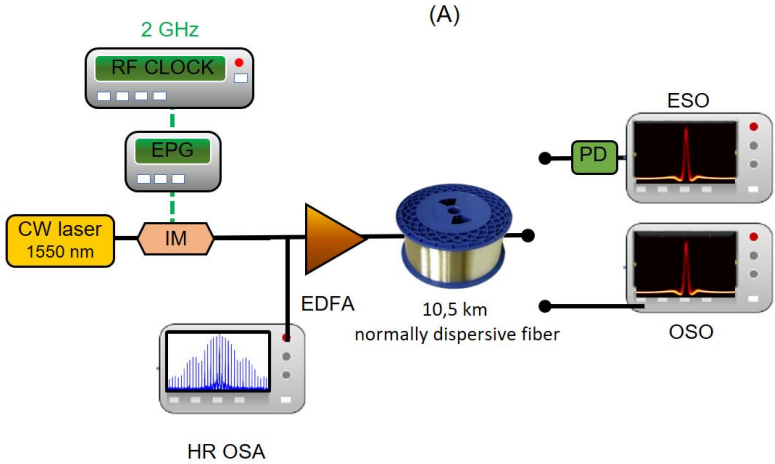
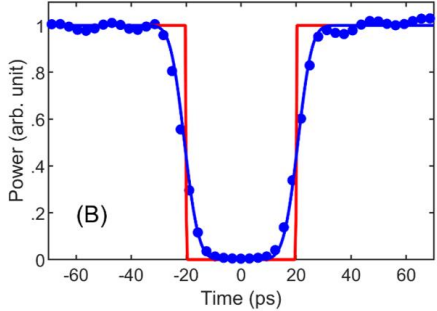
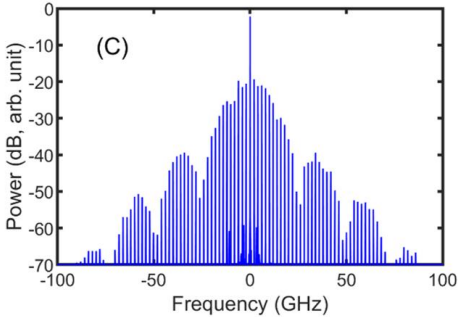

Figure 2

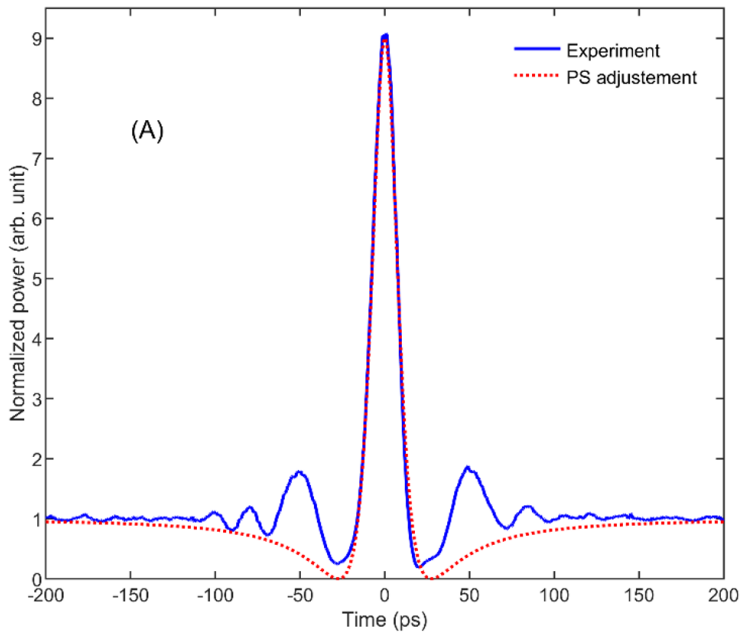
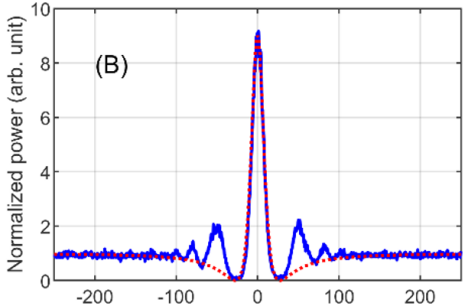
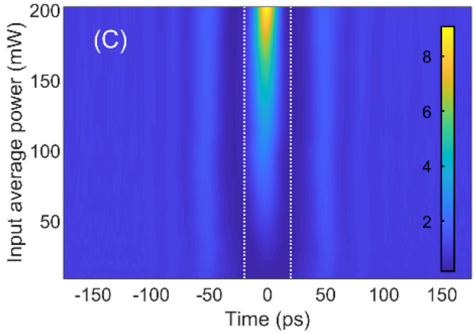

Figure 3

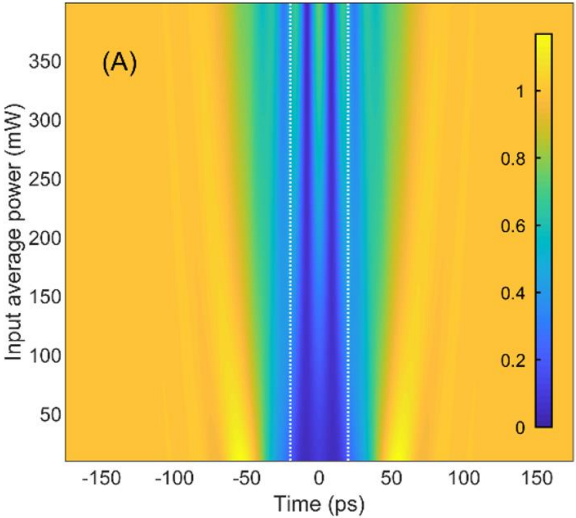 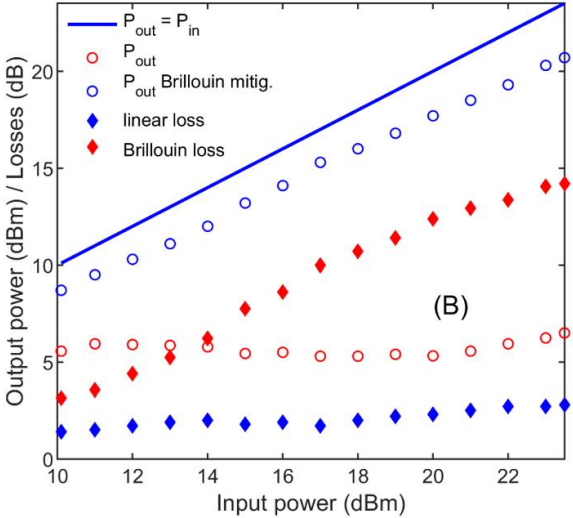

Figure 4

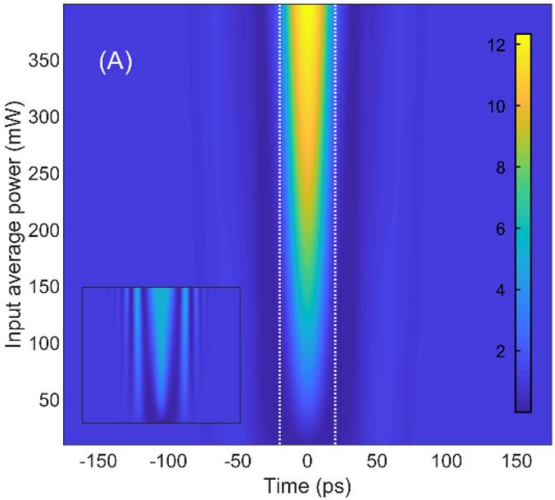 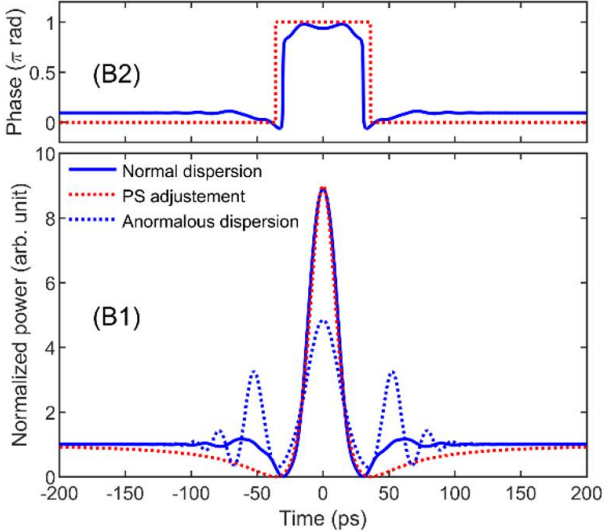